\documentclass[12pt,a4letter]{ronbun}

\usepackage{color,amsmath,amssymb,graphicx}

\usepackage{cite}
\usepackage{bm}
\usepackage{dcolumn}

\newcommand{\s}{\scriptscriptstyle}

\newcommand{\pd}{\partial}

\newcommand{\nn}{\nonumber}
\newcommand{\e}{{\rm e}}

\newcommand{\del}{\delta}

\newcommand{\al}{\alpha}

\renewcommand{\th}{\theta}

\newcommand{\ba}{\begin{eqnarray*}}
\newcommand{\ea}{\end{eqnarray*}}

\newcommand{\tg}{\tilde{\gamma}}
\newcommand{\tPi}{\widetilde{\Pi}}

\numberwithin{equation}{section}

\begin{document}

\begin{flushright}

\parbox{3.2cm}{
{KUCP-0215 \hfill \\
{\tt hep-th/0208029}}\\
\date
 }
\end{flushright}

\vspace*{0.1cm}

\begin{center}
 \Large\bf Type IIA String and Matrix String on PP-wave
\end{center}

\vspace*{0.6cm}

\centerline{\large Katsuyuki Sugiyama$^{\ast}$ 
and Kentaroh Yoshida$^{\dagger}$}

\begin{center}
$^{\ast}$\emph{Department of Fundamental Sciences, \\
Faculty of Integrated Human Studies, \\
Kyoto University, Kyoto, 606-8501, Japan.} \\
{\tt E-mail:~sugiyama@phys.h.kyoto-u.ac.jp}\\
\vspace{0.2cm}
$^{\dagger}$\emph{Graduate School of Human and Environmental Studies,
\\ Kyoto University, Kyoto 606-8501, Japan.} \\
{\tt E-mail:~yoshida@phys.h.kyoto-u.ac.jp}
\end{center}

\vspace*{1.0cm}

\centerline{\bf Abstract}

We study type IIA string theories on the pp-waves
with 24 supercharges.  
The type IIA pp-wave backgrounds are derived 
from the maximally supersymmetric pp-wave 
solution in eleven dimensions through the toroidal compactification 
on the spatial isometry directions. 
The associated actions of type IIA strings are obtained 
by using these metrics and other background fields 
of the type IIA supergravities on the one hand. On the other hand, we 
derive these theories  
from D=11 supermembrane on the pp-wave via double dimensional reduction
for the spatial isometry directions. The resulting actions agree with 
those of type IIA strings obtained in the study of the supergravities. Also, 
the action of the matrix string is written down. 
Moreover, the quantization of closed and open strings is discussed. 
In particular, we study D$p$-branes allowed in one of the type IIA
theories.  

\vspace*{1.5cm}

\thispagestyle{empty}
\setcounter{page}{0}

\newpage 

\tableofcontents

\section{Introduction}

The maximally supersymmetric 
pp-wave background in eleven dimensions 
is a classical solution (Kowalski-Glikman (KG) solution) \cite{KG} 
of the eleven-dimensional 
supergravity and is considered as one of the candidates for 
supersymmetric background of M-theory \cite{OP}.  
This pp-wave background is obtained from $AdS_7\times S^4$ or 
$AdS_4\times S^7$ via Penrose limit \cite{Penrose}. 
Recently, the maximally supersymmetric type IIB pp-wave \cite {OP2} 
has been found and 
it has been shown that the type IIB string on this pp-wave  
is exactly solvable in the Green-Schwarz (GS) 
formulation \cite{M,MT,RT} with a light-cone gauge.  
This pp-wave background \cite{OP2} is also obtained from the $AdS_5\times
S^5$ via Penrose limit \cite{Penrose}.
With this progress, the intensive studies of 
strings on the pp-waves are initiated. In particular, 
this type IIB string has been 
combined with the $AdS$/CFT correspondence 
and the almost BPS sector of a large N 
gauge theory has been studied \cite{Malda}. 
Moreover, the matrix model on the KG background has been proposed 
\cite{Malda}. 
This model is often referred as the Berenstein-Maldacena-Nastase (BMN)
matrix model. As the 
de Wit-Hoppe-Nicolai (dWHN) supermembrane 
\cite{sezgin, bergshoeff, dWHN} 
is closely related to the Banks-Fischler-Shenker-Susskind 
(BFSS) matrix model \cite{BFSS} 
in the flat space, the BMN 
matrix model is also intimately related to a supermembrane 
on the pp-wave \cite{DSR,SY,SY2}. 
In our previous works \cite{SY,SY2},  
we have shown that the algebra of supercharges in  
the supermembrane theory on the pp-wave  
agrees with that of the BMN matrix model 
in the same manner as the flat space \cite{BSS}.  We have also discussed 
BPS conditions in the supermembrane on the pp-wave.  
BPS multiplets in the BMN matrix model are also widely studied \cite{DSR,
BPS}. Moreover, the classical solutions of 
the BMN matrix and the supermembrane 
are intensively researched \cite{bak,SY3}. In particular, 
we have lately investigated the quantum stability of giant gravitons 
\cite{SY3}, 
which are classical solutions of the BMN matrix model 
and exist due to the presence of the constant 4-form flux \cite{Myers}.   

\begin{table}
\begin{center}
 \begin{tabular}{|c|ccccccccc|}
\hline
SUGRA   & 16 & 18 & 20 & 22 & 24 & 26 & 28 & 30 & 32  \\
\hline \hline 
 11 dim & $\circ$ & $\circ$ & $\circ$ & $\circ$ 
& $\circ$ & $\circ$ & $\times$ &
  $\times$ 
& unique    \\
type IIA & $\circ$  & $\circ$ & $\circ$ & $\circ$ & $\circ$ & $\circ^{\ast}$  
 & $\times$ & $\times$ &     \\
type IIB & $\circ$ & $\circ$ & $\circ$ &    & $\circ$ &    & $\circ$   
&    & unique    \\
\hline 
 \end{tabular}
\label{table}
\caption{maximal and less supersymmetric PP-waves:~ 
\footnotesize The less
 supersymmetric pp-waves are obtained by compactifications 
of the maximal pp-waves and the T-duality. The circles indicate the 
known solutions, $\times$'s that no such solutions exist, and blank 
that it is not yet known whether such solutions exist. 
The superscript 
``$\ast$'' denotes there are no
 supersymmetric D-branes.}
\end{center}
\end{table}

With recent progress, less supersymmetric type IIB and IIA 
pp-wave backgrounds, or strings
on these pp-waves are greatly 
focused \cite{twist,Pope,
GH,Warner,Gimon, Roiban, 26, Ali}. The 
maximal and less supersymmetric pp-wave backgrounds of the
eleven-dimensional supergravity, type IIA and IIB theories
are listed in Table\,\ref{table} as far as we know. 
Motivated by these attempts, 
we consider the type IIA strings  
on the pp-waves from two viewpoints in this paper.
On the one hand, 
we study the type IIA pp-waves and strings from the supergravity
side through the toroidal compactification. On the other hand, we use the 
double dimensional reduction (DDR) \cite{DDR} for the supermembrane
action on the maximally supersymmetric pp-wave. Both results are
equivalent as expected. We show that   
both compactifications 
are done for a spatial isometry direction, which can be found 
in the same way as in the type IIB case \cite{twist}.  
When we compactify this spatial direction, 
8 supercharges are inevitably broken. Therefore, 
the resulting type IIA theory has 24 supercharges, not maximally
supersymmetric. The type IIA string on this pp-wave 
is also exactly-solvable but it is different from
the one obtained from a type IIB string 
theory via the T-duality \cite{Ali}. This comes from the fact that the 
type IIA pp-wave with 24 supercharges is not unique and the 
type IIA pp-wave considered in this paper is different from 
the one in \cite{Ali}. Moreover, 
the matrix string theory is considered. 
We also discuss the quantization of closed and 
open strings in our type IIA theory. There we study the 
allowed D$p$-branes in the theory. 
The values $p=2,\,4,\,6,\,8$ are allowed but 
the directions of D-branes are restricted as in the case of type IIB
string on the pp-wave \cite{DP,BP,BGG}. 


This paper is organized as follows. 
In section 2 we consider the type IIA pp-wave backgrounds and 
actions of strings from two viewpoints. One is based on the analysis in 
the supergravity 
and the other is based on the double dimensional reduction. We will show 
both results are equivalent. In section 3 we consider the matrix 
string on the pp-wave and formally write down 
the action of the matrix string from the supermembrane action on the
pp-wave in eleven dimensions.   
In section 4 we will discuss the
mode-expansions and quantization of closed and open strings in the type IIA
theory. We also discuss D$p$-branes and investigate 
the allowed value $p$ and the  
direction of D-branes. Section 5 
is devoted to conclusions and discussions. 
In Appendix we will briefly explain the 
compactification on an $SO(3)$-direction. 
The different points from the $SO(6)$-case
considered in the text are summarized.

\section{Type IIA Strings on PP-wave} 
\label{2}

\subsection{Type IIA PP-wave Solution from KG Solution}

We can consider the toroidal compactification of a spatial isometry 
direction in 
the eleven-dimensional 
maximally supersymmetric pp-wave (KG solution) given by 
\begin{eqnarray}
ds^2 &=& - 2 dX^+ dX^- + G_{++}(X^{\s I},X^{\s I'})
(dX^+)^2 + \sum_{r =1}^9(dX^{r})^2\, , \\
&& G_{++}(X^{\s I},X^{\s I'}) \;\equiv\; - \left[\left(\frac{\mu}{3}\right)^2 
\sum_{{\s I}=1}^3(X^{\s I})^2 + \left(\frac{\mu}{6}\right)^2 
\sum_{{\s I'}=4}^9(X^{\s I'})^2 
\right]
\, , \nn  
\end{eqnarray}
where the constant 4-form flux for $+$, 1, 2, 3 directions,    
\begin{equation}
 F_{+123} \;=\; \mu\, ,\quad (\mu \neq 0) 
\label{flux}
\end{equation}
is equipped. It is a unique pp-wave solution with 32 supercharges 
in eleven dimensions. The Killing vectors of the KG solution 
are constructed as follows \cite{OP}: 
\begin{eqnarray}
 \xi_{e_+} &=& - \partial_+\,,\quad \xi_{e_-} \;=\; \partial_-\,, \nn \\
 \xi_{e_{\s I}} &=& -\cos\left(\frac{\mu}{3}X^+\right)\partial_{\s I} 
+ \frac{\mu}{3}X^{\s I}\sin\left(\frac{\mu}{3}X^+\right)\partial_-\,,
\qquad (I = 1,2,3) \,,\nn \\
 \xi_{e_{\s I}^{\ast}} &=& - \frac{\mu}{3}\sin\left(\frac{\mu}{3}X^+\right)
\partial_{\s I} - \left(\frac{\mu}{3}\right)^2 X^{\s I}
\cos\left(\frac{\mu}{3}X^+\right)\partial_{-}\,, \nn \\
 \xi_{e_{\s I'}} &=& -\cos\left(\frac{\mu}{6}X^+\right)\partial_{\s I'} 
+ \frac{\mu}{6}X^{\s I'}\sin\left(\frac{\mu}{6}X^+\right)\partial_-\,,
\qquad (I' = 4,\ldots,9)\,, \nn \\
 \xi_{e_{\s I'}^{\ast}} &=& - \frac{\mu}{6}\sin\left(\frac{\mu}{6}X^+\right)
\partial_{\s I'} - \left(\frac{\mu}{6}\right)^2X^{\s I'}
\cos\left(\frac{\mu}{6}X^+\right)\partial_{-}\,, \nn \\
\xi_{\s M_{\s IJ}} &=& X^{\s I}\partial_{\s J} - X^{\s J}\partial_{\s I}\,, 
\qquad (I,J = 1,2,3)\,, \nn \\
\xi_{\s M_{\s I'J'}} &=& X^{\s I'}\partial_{\s J'} 
- X^{\s J'}\partial_{\s I'}\,, 
\qquad (I',J' = 4,\ldots,9). \nn
\end{eqnarray}
We utilize the procedure used in 
deriving the type IIA pp-wave from the maximally supersymmetric type IIB
pp-wave through the T-duality \cite{twist}. 
The spatial isometries are given by 
\[
 \xi_{e_{\s I}} \pm \frac{3}{\mu}\xi_{e^{\ast}_{\s J}} \quad {\rm and}
 \quad 
 \xi_{e_{\s I'}} \pm \frac{6}{\mu}\xi_{e^{\ast}_{\s J'}}\,.
\]
It is sufficient to consider only two cases;
$\xi_{e_{1}} + (3/\mu)\xi_{e^{\ast}_2}$ ($SO(3)$-direction)
and 
$\xi_{e_{4}} + (6/\mu)\xi_{e^{\ast}_5}$ ($SO(6)$-direction)
due to the $SO(3)\times SO(6)$ symmetry of the KG background. 
The resulting type IIA pp-wave background has 
24 supercharges since 8 supercharges are inevitably broken 
in the toroidal compactification on the spatial isometry direction 
in the same way as the construction of type IIA pp-wave from type IIB 
via T-duality.  
In below, we will discuss mainly the $SO(6)$-direction case. 
The case of $SO(3)$ is discussed in Appendix \ref{app},  
since the story is very similar to the $SO(6)$ case though there are 
a few differences, such as the field contents.  

Let us discuss the Killing spinor in the above type IIA pp-wave. 
The Killing spinor in the KG solution is given by \cite{OP} 
\begin{eqnarray}
\epsilon(\psi_+,\,\psi_-) &=& 
\exp\left(- \frac{\mu}{4}X^+ I \right)\psi_- + \exp\left(-\frac{\mu}{12}X^+ I
\right) \psi_+ \nn \\
& & \quad + \frac{\mu}{6}\left[
\sum_{{\s I}=1}^3X^{\s I}\Gamma_{\s I} - \frac{1}{2}\sum_{{\s I'}=4}^9 
X^{\s I'}\Gamma_{\s I'}
\right] I 
\exp\left(+\frac{\mu}{12}X^+ I\right) \Gamma_- \psi_+ \, ,
\end{eqnarray}
where $\Gamma_{\mu}$'s are 32 $\times$ 32 gamma matrices and 
$I \equiv \Gamma_{123}$ obeys $I^2 = -1$. The spinors 
$\psi_+$ and $\psi_-$ with 32 components satisfy the conditions 
\begin{equation}
 \Gamma_+ \psi_+ \;=\;0\,, \quad \Gamma_- \psi_- \;=\; 0\,,
\end{equation}
hence they have 16 non-vanishing components. If $X$ is a Killing vector, 
we can define an associated 
Lie derivative $\mathcal{L}_X$ on any spinor $\psi$ by 
\begin{eqnarray}
 \mathcal{L}_X\psi &=& X^M \nabla_M\psi + \frac{1}{4}\nabla_{[M}X_{N]}\Gamma^{MN}\psi\,,
\end{eqnarray} 
where $\nabla$ is defined by 
\[
 \nabla_{M} \;\equiv\; \partial_{M}
+ \frac{1}{4}\omega_M^{ab}\Gamma_{ab}\,.
\]
This has the following properties: 
\begin{enumerate}
 \item If $X$ is a Killing vector field, $f$ is any smooth function 
and $\psi$ is any spinor, then  
\[
 \mathcal{L}_X(f\psi) \;=\; (Xf)\psi + f\mathcal{L}_X\psi\,.	
\]
\item When the symbol ``$\cdot$'' (dot) denotes the Clifford action
 of vector fields on spinors, then 
\[
 \mathcal{L}_X(Y\cdot\psi) = [X,\,Y]\cdot\psi + Y\cdot \mathcal{L}_X\psi\,.
\] 
\item If $X,\,Y$ are Killing vector fields and $\psi$ is any spinor,
       then 
\[
 \mathcal{L}_X \mathcal{L}_Y \psi - \mathcal{L}_Y\mathcal{L}_X\psi \;=\; \mathcal{L}_{[X,\,Y]}\psi\,.
\]
\end{enumerate}
The Lie derivatives for the Killing vector fields are given by \cite{OP} 
\begin{eqnarray}
&& \mathcal{L}_{\xi_{e_-}}\epsilon(\psi_+,\,\psi_-) \;=\; 0\,,\quad 
\mathcal{L}_{\xi_{e_+}}\epsilon(\psi_+,\,\psi_-) \;=\; 
\epsilon\left( - \frac{\mu}{12}I\psi_+,\, -\frac{\mu}{4}I\psi_-\right)\,, 
\nn \\
&& \mathcal{L}_{\xi_{e_{\s I}}}\epsilon(\psi_+,\,\psi_-) \;=\; 
\epsilon\left(-\frac{\mu}{6}I\Gamma_{\s I}\Gamma_-\psi_+,\, 0\right)\,, \quad 
\mathcal{L}_{\xi_{e_{\s I'}}} \epsilon(\psi_+,\,\psi_-) \;=\; 
\epsilon\left(-\frac{\mu}{12}I\Gamma_{\s I'}\Gamma_-\psi_+,\, 0\right)\,,
\nn \\ 
&& \mathcal{L}_{\xi_{e_{\s I}^{\ast}}}\epsilon(\psi_+,\,\psi_-) \;=\; 
\epsilon\left(-\frac{\mu^2}{18}
\Gamma_{\s I}\Gamma_-\psi_+,\, 0\right)\,, \quad 
\mathcal{L}_{\xi_{e_{\s I'}^{\ast}}} \epsilon(\psi_+,\,\psi_-) \;=\; 
\epsilon\left(-\frac{\mu^2}{72}
\Gamma_{\s I'}\Gamma_-\psi_+,\, 0\right)\,,
\nn \\ 
&& \mathcal{L}_{\xi_{\s M_{\s IJ}}}\epsilon(\psi_+,\,\psi_-) \;=\; 
\epsilon\left(-\frac{1}{2}\Gamma_{\s IJ}\psi_+,\, 0\right)\,, \quad 
\mathcal{L}_{\xi_{\s M_{\s I'J'}}}\epsilon(\psi_+,\,\psi_-) \;=\; 
\epsilon\left(-\frac{1}{2}\Gamma_{\s I'J'}\psi_+,\, 0\right)\,. \nn
\end{eqnarray}
By the use of the above results, we can count the remaining unbroken 
supersymmetries. For an example, in the 
case of $\xi_{e_{\s I}} + (3/\mu)\xi_{e_{\s J}^{\ast}}$, we obtain the 
following expression 
\begin{eqnarray}
\mathcal{L}_{\xi_{e_{\s I}}+ (3/\mu)\xi_{e_{\s J}^{\ast}}} 
&=&  \epsilon\left(-\frac{\mu}{3}Q\Gamma_{\s J}\Gamma_-\psi_+,\,0
\right)\,, \nn \\ 
Q^{\s IJ} &\equiv& \frac{1}{2}(1 +I\Gamma_{\s I}\Gamma_{\s J})\,. 
\end{eqnarray}  
Clearly, 16 spinors are annihilated by $\Gamma_-$. Furthermore, 
the constant matrix $Q$ plays the role of the projection operator 
and so 
annihilates additional 8 spinors in the same manner as the type IIB 
string \cite{twist}. For another example, in the case of 
$\xi_{e_{\s I'}} + (6/\mu)\xi_{e_{\s J'}^{\ast}}$, the Lie derivative 
is given by 
\begin{eqnarray}
\mathcal{L}_{\xi_{e_{\s I'}}+ (6/\mu)\xi_{e_{\s J'}^{\ast}}} 
&=&  \epsilon\left(-\frac{\mu}{6}Q\Gamma_{\s J'}\Gamma_-\psi_+,\,0
\right)\,, \nn \\ 
Q^{\s I'J'} &\equiv& \frac{1}{2}(1 + I\Gamma_{\s I'}\Gamma_{\s J'})\,. 
\end{eqnarray}  
In the same way as the case of 
$\xi_{e_{\s I}} + (3/\mu)\xi_{e_{\s J}^{\ast}}$, 24 supersymmetries are 
preserved. 
In conclusion, the above two cases of 
the type IIA pp-wave backgrounds preserve 
24 supersymmetries.

\subsection{Type IIA String from PP-wave Solution of D=11 Supergravity}

Here, we shall consider the toroidal compactification 
on an $SO(6)$ direction.
Let us transform the variables $X^r,\,(r=+,\,-,\,1,\,\ldots,\,9)$ into 
$x$'s
\begin{eqnarray}
&& X^+ \;=\; x^+\,,\quad X^- \;=\; x^- + \frac{\mu}{6}x^4x^5\,, \nn \\
&& 
X^{\s I} \;=\; x^{\s I}\,,~({I}=1,2,3) \quad X^{a} \;=\; x^a\,,~(a = 6,7,8,9)\,
 \\
&& X^4 \;=\; x^4 \cos\left(\frac{\mu}{6}x^+\right) 
- x^5\sin\left(\frac{\mu}{6}x^+\right)\,, 
\quad X^5 \;=\; x^4 \sin\left(\frac{\mu}{6}x^+\right) 
+ x^5 \cos\left(\frac{\mu}{6}x^+\right)\,, \nn
\end{eqnarray} 
then the metric is rewritten as 
\begin{eqnarray}
\label{11-trans}
 ds^2 &=& -2dx^+dx^- + G'_{++}(x^{\s I},x^a)(dx^+)^2 + \sum_{r=1}^9(dx^r)^2 
- \frac{2}{3}\mu \,x^5 dx^+ dx^4\,,  \\
&& G'_{++}(x^{\s I},x^a) \;\equiv\; -\left[
\left(\frac{\mu}{3}\right)^2
\sum_{{\s I}=1}^3(x^{\s I})^2 + \left(\frac{\mu}{6}\right)^2 
\sum_{a=6}^9(x^a)^2 \right]\,, \nn
\end{eqnarray}
but the constant 4-form flux is still expressed in Eq.(\ref{flux}). 
We can easily read off from the above metric (\ref{11-trans}) 
that the $x^4$-direction is 
a manifest spatial isometry direction \cite{twist} and obtain 
the metric of the type IIA by 
the standard technique of the dimensional reduction from 
the eleven-dimensional supergravity to the  
type IIA supergravity in ten dimensions,   
\begin{eqnarray}
 ds_{11}^2 &=& \e^{-\frac{2}{3}\phi}g_{\mu\nu}dx^{\mu}dx^{\nu} 
+ \e^{\frac{4}{3}\phi}(dy + dx^{\mu}A_{\mu})^2\,,
\label{11dim}
\end{eqnarray}
where $g_{\mu\nu}$ is a ten-dimensional metric, $A_{\mu}$ is a Kaluza-Klein 
gauge field (R-R 1-form)
and $\phi$ is a dilaton. 
The ten-dimensional metric $g_{\mu\nu}$ is given by 
\begin{eqnarray}
g_{\mu\nu}dx^{\mu}dx^{\nu} &=& 
-2dx^{+}dx^- + g_{++}(x^a,x^b)(dx^{+})^2 + \sum_{a=1}^4 (dx^a)^2 
+ \sum_{b=5}^8(dx^b)^2\,,  \\
&& g_{++}(x^a,x^b) \;\equiv\; 
- \left[\left(\frac{\mu}{3}\right)^2 \sum_{a=1}^4(x^a)^2  
+ \left(\frac{\mu}{6}\right)^2 \sum_{b=5}^8(x^b)^2
\right]\,, \nn 
\end{eqnarray}
where we have made rearrangement of 
8 coordinates $x^{1},\,x^{2},\,x^{3},\,
x^5,\,\cdots,\,x^9$ into $x^{1},\,\cdots,\,x^8$. 
The Kaluza-Klein gauge field $A_{\mu}$ is expressed as
\begin{equation}
 A_{+} \;=\; - \frac{\mu}{3}x^4\,,\quad A_{i} \;=\; 0\,,~(i=1,\ldots,8)\,,
\end{equation}
and the R-R 3-form $C_{\mu\nu\rho}$ is given by 
\begin{eqnarray}
 C_{+ {\s IJ}} \;=\; - \frac{\mu}{3}\epsilon_{\s IJK}x^{\s K}\,,~
({I,J,K} = 1,2,3)\,.
\end{eqnarray}
The dilaton $\phi$ and NS-NS 2-form $B_{\mu\nu}$ vanish.

Next, we discuss the action of the type IIA string theory 
on the above pp-wave. In below, we shall construct the action of 
bosonic and fermionic sectors, respectively. 

\noindent
\underline{\bf\sf Bosonic Sector}

In general, the bosonic world-sheet action with non-zero NS-NS B-field 
is written as  
\begin{eqnarray}
 S_{\rm B} &=& - \frac{1}{4\pi\al'}\int\! d\tau\!\! 
\int^{2\pi L}_0\!\!\!d\sigma\, \Big[g_{\mu\nu}\partial_a x^{\mu}
\partial_b x^{\nu}\eta^{ab} + B_{\mu\nu}\partial_a x^{\mu}\partial_{b}x^{\nu}
\epsilon^{ab}
\Big]\,,
\label{nlsm}
\end{eqnarray}
where  the subscript $a$ denotes the coordinates of the string
world-sheet $\sigma^a = (\tau,\sigma)$ and 
$\eta = \mbox{diag}(-1,1)$ is the world-sheet metric. 
The $L$ is the arbitrary length parameter. 
The convention of the anti-symmetric tensor is taken as
$\epsilon^{\tau\sigma}=1$. 
The ten-dimensional metric obtained previously 
and the light-cone gauge condition $x^+=\tau$ 
lead to the bosonic action of the type IIA string
theory written as 
\begin{eqnarray}
 S_{\rm B} &=& \frac{1}{4\pi \al'}\int\!d\tau\!\!\int^{2\pi}_0\!\!\! d\sigma\, 
\Biggl[\,
\sum_{i=1}^8 \left[(\partial_{\tau}x^i)^2 - 
\left(\partial_{\sigma}x^i\right)^2 \right] \nn \\ 
&& \qquad - \left(\frac{\mu}{3}\right)^2
\sum_{a=1}^4(x^a)^2 - \left(\frac{\mu}{6}\right)^2\sum_{b=5}^8(x^b)^2 
\Biggr]
\,,
\end{eqnarray}
where we have rescaled the parameters $\tau$, $\sigma$ and $\mu$ 
as 
\begin{eqnarray}
\tau \;\rightarrow\; L\tau\,,\quad \sigma \;\rightarrow\; L\sigma\,, 
\quad \mu \;\rightarrow\; \frac{\mu}{L}\,. \nn 
\end{eqnarray}

\noindent
\underline{\bf\sf Fermionic Sector}

In order to construct the fermionic action, we need explicit
expressions of the covariant derivatives. 
In the celebrated work \cite{M}, 
the covariant derivatives are obtained by the coset construction in 
$AdS_5 \times S^5$.  
However, the resulting covariant derivatives become those 
appearing in the type IIB supergravity with non-trivial 
background fluxes  
when we take the light-cone gauge conditions. Hence, we could have a 
short-cut in our hand \cite{MT}. It would be sufficient to use 
the covariant derivatives in supergravity with background fluxes and the   
expressions of the covariant derivatives are already known. 
Following the work \cite{Ali}, we can derive the 
fermionic part of the 
action using the generalized covariant derivative $\mathcal{D}$ 
defined by \footnote{It has been reported that the numerical
coefficients in the covariant derivatives in the type IIA and the type IIB 
include some issues \cite{Warner,Ali}. But it should be remarked 
that these are based on the difference of the 
convention in Ref.\,\cite{Cov}, and not on the incorrectness. 
We thank C.N.~Pope for the valuable comment on this point.} 
\begin{eqnarray}
 (\mathcal{D}_a)_{pq} &=& \partial_a \,\del_{pq} 
+ \frac{1}{4}\partial_a x^{\rho} 
\Biggl[
(\omega_{\mu\nu\rho}\,\del_{pq} 
- \frac{1}{2}H_{\mu\nu\rho}(\sigma_3)_{pq})\gamma^{\mu\nu} \nn \\
&& \qquad + \left(
\frac{1}{2\cdot 2!}F_{\mu\nu}\gamma^{\mu\nu}(i\sigma_2)_{pq}
+ \frac{1}{2\cdot 4!}F_{\mu\nu\lambda\delta}\gamma^{\mu\nu\lambda\delta}
(\sigma_1)_{pq}\right)
\gamma_{\rho}
\Biggr]\,,
\end{eqnarray}
where $\sigma_{i}$'s $(i=1,2,3)$ are Pauli matrices. The $H_{\mu\nu\rho}$ 
is the 3-form field strength of the 
NS-NS B-field. The $F_{\mu\nu}$ and $F_{\mu\nu\lambda\delta}$ are  
the 2 and 4-form field strengths of the R-R 1-form $A_{\mu}$ and 
3-form $C_{\mu\nu\lambda}$, respectively.  
Here, we have ignored the contribution of higher Kaluza-Klein modes which has 
the spectrum tower with the energy difference $1/R$. Now, the energy
difference is so large that we can ignore the $n\neq 0$ sectors. 
After this truncation, we are restricted to the $n=0$ sector and 
the gauge coupling constant $n/R$ 
is effectively zero.

Using this covariant derivative, we can obtain the quadratic fermionic 
action of the type IIA described by 
\begin{eqnarray}
S_{\rm F} &=& 
\frac{i}{2\pi}\int\!d\tau\!\!\int^{2\pi L}_0\!\!\! d\sigma\, 
\sum_{p,q,r=1}^2\left(
\eta^{ab}\del_{pq} - \epsilon^{ab}(\sigma_3)_{pq}
\right)\partial_a x^{\mu}\bar{\th}^p\gamma_{\mu}(\mathcal{D}_b)_{qr}
\th^r \,,
\end{eqnarray}
where $\th^p$'s $(p=1,2)$ are two 16-component spinors with different
chiralities in ten dimensions. When we set the light-cone gauge 
conditions, 
\[
 x^+ = \tau\,,\quad \gamma^{+}\th^p = 0\,,
\] 
then in the same way as the type IIB case \cite{MT} 
the above action can be rewritten as 
\begin{eqnarray}
 S_{\rm F} &=& -\frac{i}{2\pi}\int\!d\tau\!\!\int^{2\pi L}_0\!\!\!d\sigma\, 
\sum_{p,q,r=1}^2\bar{\th}^p \gamma_+ \left(
\del_{pq}(\mathcal{D}_{\tau})_{qr} + (\sigma_3)_{pq}
(\mathcal{D}_{\sigma})_{qr}
\right)\th^r \,, 
\end{eqnarray}
where the length parameter should be fixed as $L = \al'|p^+|$ now. 
The covariant derivatives are also rewritten as  
\begin{eqnarray}
 (\mathcal{D}_{\tau})_{pq} &=& 
\partial_{\tau}\,\del_{pq} + \frac{1}{4}\Biggl[
(\omega_{\mu\nu +}\,\del_{pq} - \frac{1}{2}H_{\mu\nu +}(\sigma_3)_{pq})
\gamma^{\mu\nu} \nn \\ 
&& \qquad 
+ \left(
\frac{1}{2\cdot 2!}F_{\mu\nu}\gamma^{\mu\nu}
+\frac{1}{2\cdot 4!}F_{\mu\nu\lambda\delta}\gamma^{\mu\nu\lambda\delta}
(\sigma_1)_{pq}
\right)
\gamma_+
\Biggr]\,, \nn \\
(\mathcal{D}_{\sigma})_{pq} &=& \partial_{\sigma}\,\del_{pq}\,.
\end{eqnarray}
When we use the constant 2 and 4-form field strengths 
$F_{+4} = \frac{\mu}{3}$ and $F_{+123} = \mu$, 
the fermionic action 
can be rewritten as  
\begin{eqnarray}
S_{\rm F} &=& \frac{i}{2\pi}\int\!d\tau\!\!\int^{2\pi}_0\!\!\!d\sigma\, 
\Biggl[\psi^{\s T}\partial_{\tau}\psi 
+ \psi^{\s T} \gamma_9 \partial_{\sigma}\psi 
+ \frac{\mu}{4}\psi^{\s T}
\left(\gamma_{123} + \frac{1}{3}\gamma_{49}
\right)
\psi
\Biggr]
\,,
\end{eqnarray}
where we have introduced a 16 component spinor $\psi$ defined by 
\[
 \psi = \dbinom{(\psi^1 + \psi^2)/\sqrt{2}}{(\psi^1 - \psi^2)/
\sqrt{2}} \;\equiv\; \dbinom{\Psi^1}{\Psi^2}
\,.
\]
The 8 component spinors $\psi^1$ and $\psi^2$ are given by 
\[
 \th^i \equiv \frac{1}{2^{1/4}}\dbinom{0}{\psi^i}\,, \quad (i=1,2)
\]
due to the light-cone conditions. Also, in the same way as in the 
bosonic sector, we have rescaled 
the parameters $\tau$, $\sigma$, $\mu$ and fermion $\psi$ as 
\begin{eqnarray}
 \tau \;\rightarrow\; L\tau\,, \quad \sigma \;\rightarrow\; 
L\sigma\,,\quad \mu \;\rightarrow\; \frac{\mu}{L}\,, \quad 
\psi \;\rightarrow\; \frac{\psi}{L}\,.
\end{eqnarray}

\subsection{Type IIA String via Double Dimensional Reduction}

By following the work \cite{dWHN} in the light-cone gauge
in terms of an $SO(9)$ spinor $\psi$, 
we can write down the action of the supermembrane on the pp-wave 
\cite{DSR,SY} as 
\begin{eqnarray}
\label{memb}
 {\cal S}&=& \frac{1}{\ell_{\s M}^3}\int\!d\tau\int^{2\pi L}_0\!\!\!
d\sigma 
\int^{2\pi L}_0\!\!\!d\rho \,{\cal L}\,,\\
 w^{-1}\mathcal{L} &=& \frac{1}{2} \Biggl[D_{\tau}X^r D_{\tau}X^r 
- \frac{1}{2}(\{X^r,\,X^s\})^2 - \left(\frac{\mu}{3}\right)^2
\sum_{{\s I}=1}^{3}X_{\s I}^2 - \left(\frac{\mu}{6}\right)^2 
\sum_{{\s I'}=4}^{9}X^2_{\s I'} \nn \\
&&  - \frac{\mu}{3} \sum_{{\s I,J,K}=1}^3\epsilon_{\s IJK}X^{\s K}
\{X^{\s I},\,X^{\s J}\} \Biggr]
+ i {\psi}^{\s T}\gamma^{r}\{X^r,\,\psi \} 
+ i{\psi}^{\s T} D_{\tau}\psi + i\frac{\mu}{4}\psi^{\s T}\gamma_{\s 123} 
\psi \, , \nn
\end{eqnarray}
where $(\sigma^0,\sigma^1,\sigma^2)=(\tau,\sigma,\rho)$ 
is the set of world-volume coordinates 
on the membrane and the $\{~ , ~ \}$ is a Lie bracket given by using 
an arbitrary function $w(\sigma,\rho)$ of 
world-volume spatial coordinates $\sigma^a$ ($a = 1,2$)
\[
 \{A\,, B\} \; \equiv \; \frac{\epsilon^{ab}}{w}\,
\partial_{a} A\, \partial_{b} B \, ,
\quad (\,a,b = 1,2\,) \,,
\]
with $\partial_a = \partial/\partial\sigma^a$.
This theory has the $\tau$-independent gauge symmetry called 
the area-preserving diffeomorphism (APD). It is a residual symmetry 
belonging to the reparametrization invariance of the membrane
world-volume. When we use the gauge connection $\omega$, 
the covariant derivative for this gauge symmetry
is defined by 
\begin{equation}
 D_{\tau} X^r \;\equiv\; \pd_{\tau}X^r - \{\omega,\,X^r \}\,,\qquad
(r=1,2,\cdots ,9) . \nn
\end{equation}
We have also introduced a parameter $\ell_{\s M}$, which 
is the M-theory scale 
related to the membrane tension $T_{\s M} = 1/\ell_{\s M}^3$.
It is associated to the string coupling $g_s$ 
and the string scale $\ell_s$ in ten-dimensional string theory 
(up to some numerical constant) 
with a relation $\ell_{\s M} = g_s^{1/3}\ell_s$. 
We use a normalization 
\begin{eqnarray}
0 \;\leq\; \sigma \;\leq\; 2\pi L\,,\quad 
0 \;\leq\; \rho \;\leq\;  2\pi L\,, \qquad 
\int\! d\sigma d\rho\,w(\sigma,\rho) \;=\; L^2\,,\nn
\end{eqnarray}
with $L$ being an arbitrary length parameter.
In our light-cone gauge, the time coordinate ``$\tau$'' is associated to 
the $X^{+}$ as $\displaystyle 
X^{+} = (\ell_{\s M}^3/(2\pi L)^2)P^{+}_0\tau$ and 
the longitudinal momentum $P^{+}(\sigma,\rho)$ satisfies 
$\displaystyle P^{+}(\sigma,\rho)=(P^{+}_0/L^2)\, w(\sigma,\rho)$. 
Hereafter we shall use a convention $P^{+}_0 = 1$.  

Here, we shall consider the double dimensional reduction (DDR) 
in the $SO(6)$-direction. It is considered that 
11 dimensional supermembrane theory in the flat space 
should reduce to the type IIA string theory, at least classically. 
Based on this fact, we shall carry out the DDR of the 
supermembrane on the pp-wave. We will show that 
the type IIA string action on the pp-wave obtained in the 
previous subsection  
can be derived from the supermembrane action 
on the pp-wave (\ref{memb}) 
through the double dimensional reduction.  

To begin, we rotate the variables $X^{1},\,\cdots,\,X^9$ into $x$'s
\begin{eqnarray}
&& X^{\s I} \;=\; x^{\s I}\,,\quad (I = 1,2,3), \quad 
X^a \;=\; x^a\,, \quad (a = 6,7,8,9), \nn \\
&& X^4 \;=\; x^4\cos\left(\frac{\mu}{6}\tau\right) 
- x^5\sin\left(\frac{\mu}{6}\tau\right)\,, \quad X^5 \;=\; 
x^4 \sin\left(\frac{\mu}{6}\tau\right) + x^5 \cos\left(\frac{\mu}{6}\tau
\right)\,,
\label{rotate}
\end{eqnarray}
then an associated action is written down as follows:
\begin{eqnarray}
 \mathcal{S} &=& \frac{1}{\ell_{\s M}^3}\int\! d\tau\!\!\int^{2\pi }_0
\!\!\! d\sigma\!\! \int^{2\pi L}_0\!\!\! d\rho\, \mathcal{L}\,,  \\
w^{-1} \mathcal{L} &=& \frac{1}{2}
\Biggl[
(D_{\tau}x^r)^2 - \frac{1}{2}(\{x^r,\,x^s\})^2 - \left(\frac{\mu}{3}\right)^2
\sum_{{\s I}=1}^3(x^{\s I})^2 - \left(\frac{\mu}{6}\right)^2\sum_{{\s I'}=6}^9
(x^{\s I'})^2 \nn \\
&& \qquad - \frac{\mu}{3}\sum_{{\s I,J,K}=1}^3\epsilon_{\s IJK}x^{\s K}
\{x^{\s I},\,x^{\s J}\} 
- \frac{2}{3}\mu\, x^5 D_{\tau}x^4 
\Biggr] \nn \\
&& \qquad 
+ i\psi^{\s T}\gamma^r\{x^r,\,\psi\} 
+ i \psi^{\s T}D_{\tau}\psi + i\frac{\mu}{4}\psi^{\s T}
\left(\gamma_{123} + \frac{1}{3}\gamma_{54} \right)\psi\,, \nn 
\end{eqnarray}
where 
it should be noted that the fermion mass term is modified 
compared with flat case. 
This contribution appears since we have moved to the rotated 
coordinate. In this time, the additional spin connection
$\omega_+^{4\,5} = \mu/6$ appears 
\footnote{
We have modified the contribution of the spin connection 
in the revised version. This contribution is initially 
pointed out in Ref.\,\cite{HS} where the correct type IIA action is 
obtained. }. 
Now let us consider the DDR in the $x^4$-direction. 
That is, we take $x^4 = \rho$. 
We choose the density function 
$w(\sigma,\rho)$ to be a constant so that $w=(2\pi)^{-2}$ 
and fix the parameter $L$ as $L=g_s\ell_s$. 
The resulting action is given by  
\begin{eqnarray}
S_{st} &=&\frac{1}{2\pi}\int\!d\tau\!\!\int^{2\pi}_0\!\!\!d\sigma \,
\mathcal{L}_{st}\,,\nn\\
\mathcal{L}_{st} &=& \frac{1}{2\alpha'}\left[
\sum_{i=1}^8\left[
(\partial_{\tau}x^i)^2 - (\partial_{\sigma}x^i)^2\right] 
- \left(\frac{\mu}{3}\right)^2 \sum_{{a}=1}^4 (x^{a})^2 
- \left(\frac{\mu}{6}\right)^2\sum_{b = 5}^8 (x^b)^2\right] \nn \\
&&\qquad  + i\psi^{\s T} \left[{\bf 1}_{16}\cdot\partial_{\tau} 
- \gamma_9 \cdot\partial_{\sigma} + \frac{\mu}{4}
\left(\gamma_{\s 123} + \frac{1}{3}\gamma_{49} \right)
\right]\psi \,,
\end{eqnarray}
where we have renamed the coordinate $x^4,\,x^5,\,\ldots,\,x^9$ 
as $x^9,\,x^4,\,x^5,\,\ldots,\,x^8$. It should be understood that 
the mass term of $x^4$ arises from the second term 
in Eq.\,(\ref{11dim}). This term 
should describe the effect of the Kaluza-Klein 1-form. 
Also, the fermionic field and parameters has been appropriately rescaled as 
\begin{eqnarray}
&&\sigma \;\rightarrow\;  L\,\sigma\,, \quad 
\tau \;\rightarrow\; \frac{L}{(2\pi)^2}\,\tau\,,\quad 
\psi \;\rightarrow\; \frac{\ell_{\s M}^{3/2}}{L}\,\psi\,,\quad 
\mu \;\rightarrow\; \frac{(2\pi)^2}{L}\,\mu\,.\nn
\end{eqnarray}
The parameters of the resulting theory are related 
with those of M-theory 
\begin{eqnarray}
&&\frac{1}{2\pi\alpha'} \;=\; \frac{(2\pi)L}{\ell_{\s M}^3}\,,\quad 
\ell_{\s M} \;=\; g_s^{1/3}\ell_s\,,\quad 
L \;=\; g_s\ell_s\,,\quad \ell_s \;=\; 2\pi\sqrt{\alpha'}\,.\nn
\end{eqnarray}

It should be noted that the above action is identical with 
the type IIA action 
derived in the previous subsection 
up to the sign of $\sigma$. 
Hereafter,  
we can use the following expression of $\gamma^{\mu} 
= (\gamma^i,\, \gamma^8,\,\gamma^9)$, 
\begin{eqnarray}
&& \gamma^i \;=\; \tg^{i}\otimes \sigma_2  
\;=\;  
\begin{pmatrix}
0 & -i\tg^i \\
i\tg^i & 0 
\end{pmatrix}
\,, \quad (i=1,\,\ldots,\,7)\,, \\  
&& \gamma^{8} \;=\; {\bf 1}_8 \otimes \sigma_1  
\;=\;
\begin{pmatrix}
0 & {\bf 1}_8 \\
{\bf 1}_8 & 0 
\end{pmatrix}
\,, \quad 
\gamma^9 \;=\; {\bf 1}_8 \otimes \sigma_3 \;=\;
\begin{pmatrix}
{\bf 1}_8 & 0 \\
0 & - {\bf 1}_8 
\end{pmatrix}
\,,
\end{eqnarray} 
where $\tg^i$'s $(i=1,\,\ldots,\,7)$ are $SO(7)$ gamma matrices that obey 
commutation relations 
\begin{eqnarray}
\tg^i\tg^j + \tg^j\tg^i \;=\; 2\del^{ij}\,.
\end{eqnarray}
The 16 component fermion $\psi$ is decomposed into  
two 8 component fermions $\Psi^1$ and $\Psi^2$ as 
\[
\psi \;=\; \dbinom{\Psi^1}{\Psi^2}\,.
\] 
Moreover, we can decompose the 8 component fermion into two 
eigen-spinors of the matrix $R \equiv \tg_{1234}$ as follows:
\begin{eqnarray}
\Psi^a &=& \left(\frac{1+R}{2}\right)\Psi^a 
+ \left(\frac{1-R}{2}\right)\Psi^a \nn \\
&\equiv& \Psi^{a+} + \Psi^{a-}\,,\qquad (a=1,2)\,.
\end{eqnarray}
By definition, the spinor $\Psi^{a\pm}$ 
satisfies 
\begin{equation}
R\Psi^{a\pm} \;=\; \pm\Psi^{a\pm}\,.
\end{equation} 
That is, $\Psi^{a\pm}$ is the eigen-spinor with eigen-value $\pm 1$. 
By the use of $\Psi^{a\pm}$, the fermionic Lagrangian can be rewritten as 
\begin{eqnarray}
\mathcal{L} &=& i \Psi^{1+}{}^{\s T}\partial_-\Psi^{1+} 
+ i\Psi^{1-}{}^{\s T}\partial_-\Psi^{1-} 
+ i\Psi^{2+}{}^{\s T}\partial_+\Psi^{2+} 
+ i\Psi^{2-}{}^{\s T}\partial_+\Psi^{2-} \nn \\
&& - i\frac{\mu}{3}\Psi^{1-}\widetilde{\Pi}^{\s T}\Psi^{2+} 
- i\frac{\mu}{6}\Psi^{1+}{}^{\s T}\widetilde{\Pi}^{\s T}\Psi^{2-} 
+ i\frac{\mu}{6}\Psi^{2-}{}^{\s T}\widetilde{\Pi}\Psi^{1+} 
+ i\frac{\mu}{3}\Psi^{2+}{}^{\s T}\widetilde{\Pi}\Psi^{1-}\,,
\end{eqnarray}
where $\widetilde{\Pi} \equiv \tg_{123}$,\, $\widetilde{\Pi}^{\s T}\equiv 
\tilde{\gamma}_{321}$ and satisfies 
$\widetilde{\Pi}\widetilde{\Pi}^{\s T} = 
\widetilde{\Pi}^{\s T}\widetilde{\Pi} = 1$.

\section{Matrix String Theory on PP-wave} 

We can also consider the matrix string theories
\cite{m-string} on the pp-wave \footnote{Matrix strings are also
discussed in Refs.\,\cite{Bonelli,Gopakumar} 
from different viewpoints from ours.} 
from the supermembrane by the use of the method 
in the work \cite{Yoneya}.  

Let us start with the supermembrane action (\ref{memb}), and 
rotate the variables into $x$'s as given by (\ref{rotate}). 
In this time, the gamma matrices are also transformed by this rotation.
The resulting supermembrane action is given by 
\begin{eqnarray}
 \mathcal{S} &=& \frac{1}{\ell_{\s M}^3}\int\! d\tau\!\!\int^{2\pi }_0
\!\!\! d\sigma\!\! \int^{2\pi L}_0\!\!\! d\rho\, \mathcal{L}\,,  \\
 \mathcal{L} &=& \frac{1}{2}
\Biggl[
(D_{\tau}x^r)^2 - \frac{1}{2}(\{x^r,\,x^s\})^2 - \left(\frac{\mu}{3}\right)^2
\sum_{{\s I}=1}^3(x^{\s I})^2 - \left(\frac{\mu}{6}\right)^2\sum_{{\s I'}=6}^9
(x^{\s I'})^2 \nn \\
&& \qquad - \frac{\mu}{3}\sum_{{\s I,J,K}=1}^3\epsilon_{\s IJK}x^{\s K}
\{x^{\s I},\,x^{\s J}\} 
- \frac{2}{3}\mu\, x^5 D_{\tau}x^4 
\Biggr] \nn \\
&& \qquad 
+ i\psi^{\s T}\gamma^r\{x^r,\,\psi\} 
+ i \psi^{\s T}D_{\tau}\psi + i\frac{\mu}{4}\psi^{\s T}
\left(\gamma_{123} + \frac{1}{3}\gamma_{54} \right)\psi\,, \nn 
\end{eqnarray}
where we have set $w = (2\pi)^{-2}$ and rescaled $\sigma$ as 
$\sigma \rightarrow (2\pi)^2\sigma$. Now, the Lie bracket $\{~\,,\,~\}$ is 
simply defined by 
\[
 \{A,\,B\} \;\equiv\; \partial_{\sigma}A\,\partial_{\rho}B 
- \partial_{\rho}A\,\partial_{\sigma}B\,.
\] 
Then, we rewrite $x^4$ as $x^4 \equiv Y$ and shift $Y$ as 
\[
 Y \longrightarrow \rho + Y\,.
\] 
The $Y$ is regarded as the compactified direction.
As the result, the action is rewritten as 
\begin{eqnarray}
\label{2.29}
 \mathcal{S} &=& \frac{L}{\ell_{\s M}^3}\int\! d\tau\!\!\int^{2\pi }_0
\!\!\! d\sigma\!\! \int^{2\pi }_0\!\!\! d\rho\, \mathcal{L}\,,  \\
\mathcal{L} &=& \frac{1}{2}
\Biggl[ F_{0,\sigma}^2 
+(D_{\tau}x^i)^2 - (D_{\sigma}^{\s Y}x^i)^2 
- \frac{1}{2L^2}(\{x^i,\,x^j\})^2 - \left(\frac{\mu}{3}\right)^2
\sum_{{\s I}=1}^3(x^{\s I})^2 - \left(\frac{\mu}{6}\right)^2\sum_{{\s I'}=5}^8
(x^{\s I'})^2 \nn \\
&& \qquad - \frac{\mu}{3L}\sum_{{\s I,J,K}=1}^3\epsilon_{\s IJK}x^{\s K}
\{x^{\s I},\,x^{\s J}\} 
- \frac{2}{3}\mu\, x^4 F_{0,\sigma} 
\Biggr] \nn \\
&& \qquad 
+ i\frac{1}{L}\psi^{\s T}\gamma^i\{x^i,\,\psi\} 
+ i \psi^{\s T}D_{\tau}\psi 
-i\psi^{\s T}\gamma^9 D_{\sigma}^{\s Y}\psi
+ i\frac{\mu}{4}\psi^{\s T}
\left(\gamma_{123} + \frac{1}{3}\gamma_{49} \right)\psi\,, \nn 
\end{eqnarray}
where we have reassigned the variables $x^4,\,x^5,\,\ldots,\,x^9$ as 
$x^9,\,x^4,\,x^5\,\ldots,\,x^8$ and rescaled $\rho \rightarrow
L\,\rho$. We have also introduced the following quantities, 
\begin{eqnarray}
 F_{0,\sigma} &\equiv& \partial_{\tau}Y - \partial_{\sigma}\omega 
- \frac{1}{L}\{\omega,\,Y\}\,,\nn \\
 D_{\tau}x^i &\equiv& \partial_{\tau}x^i - \frac{1}{L}\{\omega,\,x^i\}\,, 
\nn \\
 D_{\sigma}^{\s Y}x^i &\equiv& \partial_{\sigma} x^{i} - \frac{1}{L}\{Y,\,x^i\}\,, \nn 
\end{eqnarray}
where $A_0 \equiv \omega$ and $A_{\sigma} \equiv Y$. The inverse
compactification radius $1/L$ plays a role of the gauge coupling
constant. 
It seems that
the action (\ref{2.29}) is not explicitly
invariant under
the are-preserving diffeomorphism. 
But the action indeed has this symmetry 
under the transformation with an infinitesimal gauge 
parameter $\Lambda$
\begin{eqnarray*}
&&\delta\omega \;=\; L\partial_{\tau}\Lambda +\{\Lambda ,\omega\}\,,\\
&&\delta Y \;=\; L\partial_{\sigma}\Lambda +\{\Lambda ,Y\}\,,\\
&&\delta x^i \;=\; \{\Lambda ,x^i\}\,,\quad 
\delta\psi \;=\; \{\Lambda ,\psi\}\,.
\end{eqnarray*}

The action (\ref{2.29}) is very close to that of the matrix string
theory. In fact, 
using the corresponding law of Ref.\, \cite{Yoneya} in the large $N$
limit, it is straightforward to map the supermembrane action into 
matrix representations. Thus, 
we can formally obtain the matrix string action on the pp-wave, 
up to $O(1/N^2)$
described by 
\begin{eqnarray}
\label{mst}
  \mathcal{S} &=& \frac{L}{\ell_{\s M}^3}\int\! d\tau \frac{2\pi}{N}
\int^{2\pi }_0\!\!\! d\th\, \mathcal{L}\,,  \\
\mathcal{L} &=& \frac{1}{2}{\rm Tr}
\Biggl[ F_{0,\th}^2 
+(D_{\tau}x^i)^2 - N^2(D_{\th}x^i)^2 
+ \frac{1}{2}\left(\frac{N}{2\pi L}\right)^2
([x^i,\,x^j])^2 \nn \\
&& \qquad - \left(\frac{\mu}{3}\right)^2
\sum_{{\s I}=1}^3(x^{\s I})^2 - \left(\frac{\mu}{6}\right)^2\sum_{{\s I'}=5}^8
(x^{\s I'})^2 \nn \\
&& \qquad + i \frac{\mu}{3}\left(\frac{N}{2\pi L}\right)
\sum_{{\s I,J,K}=1}^3\epsilon_{\s IJK}
x^{\s K}[x^{\s I},\,x^{\s J}] 
- \frac{2}{3}\mu\, x^4 F_{0,\th} 
\Biggr] \nn \\
&& \qquad 
+ {\rm Tr}\left[\frac{N}{2\pi L}\psi^{\s T}\gamma^i [x^i,\,\psi]
+ i \psi^{\s T}D_{\tau}\psi 
- N i\psi^{\s T}\gamma^9 D_{\th}\psi
+ i\frac{\mu}{4}\psi^{\s T}\left(
\gamma_{123} + \frac{1}{3}\gamma_{49} \right)
\psi\right] \,, \nn 
\end{eqnarray}
where the quantities in the action is replaced with 
\begin{eqnarray}
 F_{0,\th} &=& \partial_{\tau}Y - N\partial_{\th}\omega 
+ i\frac{N}{2\pi L}[\omega,\,Y]\,,\nn \\
 D_{\tau}x^i &=& \partial_{\tau}x^i + i\frac{N}{2\pi L}
[\omega,\,x^i]\,, \nn \\ 
D_{\th}x^i &=& \partial_{\th} x^{i} +i \frac{1}{2\pi L}
[Y,\,x^i]\,. \nn 
\end{eqnarray}
If we rescale some constants as 
\begin{eqnarray}
\tau \rightarrow \frac{\tau}{N}\,,\quad \psi \rightarrow \sqrt{N}\,\psi\,, 
\quad L \rightarrow \frac{L}{2\pi}\,, \quad \mu \rightarrow N\,\mu\,, \nn 
\end{eqnarray}
then we can rewrite Eq.(\ref{mst}) as 
\begin{eqnarray}
\label{final}
  \mathcal{S} &=& \frac{1}{\ell_{\s M}^3}\int\! d\tau \!\!
\int^{2\pi }_0\!\!\! d\th\, \mathcal{L}\,,  \\
\mathcal{L} &=& \frac{1}{2}{\rm Tr}
\Biggl[ \left(F_{0,\th} - \frac{\mu}{3}x^4\right)^2 
+(D_{\tau}x^i)^2 - (D_{\th}x^i)^2 
+ \frac{1}{2}([x^i,\,x^j])^2 \nn \\
&& \qquad - \left(\frac{\mu}{3}\right)^2
\sum_{{\s I}=1}^4(x^{\s I})^2 - \left(\frac{\mu}{6}\right)^2\sum_{{\s I'}=5}^8
(x^{\s I'})^2 + i \frac{2}{3}\mu 
\sum_{{\s I,J,K}=1}^3\epsilon_{\s IJK}
x^{\s I}x^{\s J}x^{\s K}
\Biggr] \nn \\
&& \qquad 
+ {\rm Tr}\left[\psi^{\s T}\gamma^i [x^i,\,\psi]
+ i \psi^{\s T}D_{\tau}\psi 
- i\psi^{\s T}\gamma^9 D_{\th}\psi
+ i\frac{\mu}{4}\psi^{\s T}
\left(\gamma_{123} + \frac{1}{3}\gamma_{49} \right)
\psi\right]\,, \nn 
\end{eqnarray}
where the field strength and covariant derivatives are given by 
\begin{eqnarray}
 F_{0,\th} &=& \partial_{\tau}Y - \partial_{\th}\omega 
+ i[\omega,\,Y]\,,\nn \\
 D_{\tau}x^i &=& \partial_{\tau}x^i +i [\omega,\,x^i]\,, \nn \\ 
D_{\th}x^i &=& \partial_{\th} x^{i} +i [Y,\,x^i]\,. \nn 
\end{eqnarray}
The action (\ref{final}) includes the 3-point interaction and several 
mass terms and also the field strength of the gauge connection is 
shifted by $x^4$. Thus it seems that 
the action(\ref{final}) is not invariant 
under the area-preserving diffeomorphism. 
However this action of the matrix string is actually 
invariant under the gauge transformation 
with an matrix parameter $\Lambda$
\begin{eqnarray*}
&&\delta\omega \;=\; \partial_{\tau}\Lambda -i[\Lambda ,\omega]\,,\nn\\
&&\delta Y \;=\; \partial_{\theta}\Lambda -i[\Lambda ,Y]\,,\nn\\
&&\delta x^i \;=\; -i[\Lambda ,x^i]\,,\quad 
\delta\psi \;=\; -i[\Lambda ,\psi]\,.
\end{eqnarray*}
The $\tau$-scaling leads to the $N$ dependence of 
the physical light-cone time $X^+$ as 
\[
 X^+ \;=\; \frac{\ell_{\s M}^3\,P^+_0\,\tau}{N (2\pi L)^2} \,.
\] 
We also should rescale as $P^+_0 \rightarrow N\,P^+_0$ so that 
$X^+$ should be independent of $N$. 
The diagonal elements of the matrix $x^i$ describe a fundamental 
string bit in the large $N$ limit and hence the total longitudinal 
momentum is proportional to the number $N$ of string bits. 

It is easily observed that the above action (\ref{final}) 
becomes the usual matrix string action in the flat limit $\mu
\rightarrow 0$. 
Moreover, let us consider the IR region. At the time, 
the matrix variables are restricted to the 
Cartan subalgebra. That is, the matrix becomes diagonal and so 
the term including commutator should vanish. Finally, integrating out 
the field strength $F_{0,\th}$ as the auxially field, 
one can find that 
the above action should reduce to the 
free type IIA string theory obtianed in the previous section, 
as is expected. 

Also, we should remark that 
the above action of the matrix string is included 
in the family of the work \cite{Bonelli} where the action of the 
matrix string and supersymmetry 
have been more generally investigated from the 
viewpoint of the mass deformation of the Yang-Mills theory.

Finally, we comment on the classical solution. 
As in the BMN matrix model \cite{Malda}, for an example, 
this matrix string theory has 
the static fuzzy sphere solution described by 
\begin{eqnarray}
&& x^{\s I} \;=\; \frac{\mu}{3}J^{\s I}\,,\quad (I = 1,\,2,\,3)\,, \nn \\
&& x^{4} \;=\; \cdots \;=\; x^8 \;=\; Y \;=\; \omega 
\;=\;  0\,,  
\end{eqnarray}
where the $J^{\s I}$'s are generators of an $SU(2)$ algebra.  
The existence of the fuzzy sphere solution might be 
physically 
expected from the presence of the 
constant flux of R-R 3-form \cite{Myers}. 
It would be possible to consider other classical solutions.

\section{Quantization of Type IIA String on PP-wave}

In this section we will consider the mode-expansions and 
quantization of closed and open strings in the type IIA on the pp-wave. 
In particular, we investigate D-branes living in the theory. 

\subsection{Closed Strings in Type IIA on PP-wave}

In this subsection we will discuss the mode-expansions of closed 
bosonic and fermionic
degrees of freedom and consider 
the quantization of the type IIA string.

First the variation of the action previously obtained 
leads to equations of motion 
given by 
\begin{eqnarray}
\label{em1}
 \partial_+ \partial_- x^a + \frac{\mu^2}{9}\,x^a &=& 0\,, \quad 
(a=1,2,3,4)\,,  \\
 \partial_+ \partial_- x^b + \frac{\mu^2}{36}\,x^b &=& 0\,, \quad
(b=5,6,7,8) \,,  \\
 \partial_+ \Psi^{2+} + \frac{\mu}{3}\,\widetilde{\Pi}\,\Psi^{1-} 
&=& 0\,,  \\ 
 \partial_- \Psi^{1-} - \frac{\mu}{3}\,\widetilde{\Pi}^{\s T}\,
\Psi^{2+} &=& 0\,,  \\ 
  \partial_+ \Psi^{2-} + \frac{\mu}{6}\,\widetilde{\Pi}\,\Psi^{1+} 
&=& 0\,,  \\ 
 \partial_- \Psi^{1+} - \frac{\mu}{6}\,\widetilde{\Pi}^{\s T}\,
\Psi^{2-} &=& 0\,. \label{em6} 
\end{eqnarray}
The mode-expansions of bosonic variables are described by 
\begin{eqnarray}
x^a(\tau,\,\sigma) &=&  
x_0^{a} \cos\left(\frac{\mu}{3}\tau\right) 
+ \left(
\frac{3}{\mu}\right)\al' p_0^{a}\sin\left(\frac{\mu}{3}\tau\right) \nn \\
& & \qquad \qquad 
+ i\sqrt{\frac{\al'}{2}}\sum_{n\neq 0}\frac{1}{\omega_n^{\rm\s B}}
\left[\alpha_n^{a}\phi_n^{\s\rm B} + \bar{\alpha}_n^{a} 
\tilde{\phi}_n^{\s\rm B} \right]
\,,  
\quad (a=1,2,3,4)\,,\nn \\
x^b(\tau,\,\sigma) &=& 
x_0^{b} \cos\left(\frac{\mu}{6}\tau\right) 
+ \left(\frac{6}{\mu}\right)
\al' p_0^{b}\sin\left(\frac{\mu}{6}\tau\right) \nn \\
&& \qquad \qquad 
+ i\sqrt{\frac{\al'}{2}}\sum_{n\neq 0}\frac{1}{\omega_n^{\rm\s B'}}
\left[
\alpha_n^{b}\phi_n^{\s\rm B'} + \bar{\alpha}_n^b\tilde{\phi}_n^{\s\rm B'}  
\right]\,, 
\quad (b=5,6,7,8)\,,\nn 
\end{eqnarray}
and those of fermionic variables are 
\begin{eqnarray}
 \Psi^{1-}(\tau,\,\sigma) &=& \Psi_0 \cos\left(\frac{\mu}{3}\tau\right) 
+ \widetilde{\Psi}_0\sin\left(\frac{\mu}{3}\tau\right) \nn \\
&& \qquad + \sum_{n\neq 0}c_n\left(\frac{3}{\mu}i(\omega_n^{\rm\s F} - n)
\,\tPi^{\s T}
\Psi_n\phi_n^{\s\rm F} + \widetilde{\Psi}_n\tilde{\phi}_n^{\rm\s F}\right) 
\,, \\
\Psi^{2+}(\tau,\,\sigma) &=& - \tPi\Psi_0\sin\left(\frac{\mu}{3}\tau\right) 
+ \tPi\widetilde{\Psi}_0\cos\left(\frac{\mu}{3}\tau\right) \nn \\
&& \qquad + \sum_{n\neq 0}
c_n\left[\Psi_n\phi_n^{\s\rm F} - \frac{3}{\mu}i(\omega_n^{\rm\s F} -n)
\,\tPi\widetilde{
\Psi}_n\tilde{\phi}_n^{\s\rm F}\right] \,,  \\
 \Psi^{1+}(\tau,\,\sigma) &=& \Psi_0' \cos\left(\frac{\mu}{6}\tau\right) 
+ \widetilde{\Psi}_0'\sin\left(\frac{\mu}{6}\tau\right) \nn \\
&& \qquad + \sum_{n\neq 0}c_n'\left(\frac{6}{\mu}i(\omega_n^{\rm\s F'} - n)
\,\tPi^{\s T}
\Psi_n'\phi_n^{\s\rm F'} + \widetilde{\Psi}_n'\tilde{\phi}_n^{\rm\s F'}
\right) \,, \\
\Psi^{2-}(\tau,\,\sigma) &=& - \tPi\Psi_0'\sin\left(\frac{\mu}{6}\tau\right) 
+ \tPi\widetilde{\Psi}_0'\cos\left(\frac{\mu}{6}\tau\right) \nn \\
&& \qquad + \sum_{n\neq 0}
c_n'\left[\Psi_n'\phi_n^{\s\rm F'} - \frac{6}{\mu}i(\omega_n^{\rm\s F'} -n)
\,\tPi\widetilde{\Psi}_n'\tilde{\phi}_n^{\s\rm F'}\right] \,, 
\end{eqnarray} 
where we introduced several notations
\begin{eqnarray}
&& \omega_n^{\rm\s B} 
\;=\; {\rm sgn}(n)\sqrt{n^2 + \left(\frac{\mu}{3}\right)^2}\,,\quad 
\omega_n^{\rm\s B'} \;=\; 
{\rm sgn}(n)\sqrt{n^2 + \left(\frac{\mu}{6}\right)^2 }
\,, \label{omega} \\ 
&& \phi_n^{\s\rm B} 
\;=\; \exp\left(-i(\omega_n^{\rm\s B}\tau - n\sigma)\right)\,,\quad 
\tilde{\phi}_n^{\rm\s B} 
\;=\; \exp\left(-i\left(\omega_n^{\rm\s B}\tau  + n\sigma\right)
\right)\,, \nn \\
&& \phi_n^{\s\rm B'} 
\;=\; \exp\left(-i(\omega_n^{\rm\s B'}\tau - n\sigma)\right)\,,\quad 
\tilde{\phi}_n^{\rm\s B'} 
\;=\; \exp\left(-i\left(\omega_n^{\rm\s B'}\tau  + n\sigma\right)
\right)\,, \nn \\
&& \omega^{\rm\s F}_n \;=\; 
{\rm sgn}(n)\sqrt{n^2 + \left(\frac{\mu}{3}\right)^2 }\,, \quad 
\omega^{\rm\s F'}_n \;=\; 
{\rm sgn}(n)\sqrt{n^2 + \left(\frac{\mu}{6}\right)^2 }\,, 
\nn \\
&& \phi_n^{\s\rm F} 
\;=\; \exp\left(-i(\omega_n^{\rm\s F}\tau - n\sigma)\right)\,,\quad 
\tilde{\phi}_n^{\rm\s F} 
\;=\; \exp\left(-i\left(\omega_n^{\rm\s F}\tau  + n\sigma\right)
\right)\,, \nn \\
&& \phi_n^{\s\rm F'} 
\;=\; \exp\left(-i(\omega_n^{\rm\s F'}\tau - n\sigma)\right)\,,\quad 
\tilde{\phi}_n^{\rm\s F'} 
\;=\; \exp\left(-i\left(\omega_n^{\rm\s F'}\tau  + n\sigma\right)
\right)\,, \nn \\
&& c_n \;=\; \left(1 + \left(\frac{3}{\mu}\right)^2(
\omega_n^{\rm\s F } - n)^2\right)^{-1/2}\,, \quad 
c_n' \;=\; \left(1 + \left(\frac{6}{\mu}\right)^2(
\omega_n^{\rm\s F' } - n)^2\right)^{-1/2}
\,. \nn 
\end{eqnarray}
Following the usual canonical quantization procedure, we can quantize
the theory and 
obtain the commutation relations. The canonical momenta are given
by 
\begin{eqnarray}
p^a &=& \frac{1}{2\pi\al'}
\partial_{\tau}x^a \,,\quad (a = 1,2,3,4)\,, \nn \\
p^b &=& \frac{1}{2\pi\al'}
\partial_{\tau}x^b \,,\quad (b = 5,6,7,8)\,, \nn \\
S_{\al} &=& i\psi^{\s T}_{\al}\,, \nn 
\end{eqnarray}
and the canonical (anti-)commutation relations are represented as 
\begin{eqnarray}
[x^i(\tau,\,\sigma),\,p^j(\tau,\sigma')] &=& 
i \del^{ij}\,\del(\sigma - \sigma')\,, \nn \\
\{\psi_{\al}(\tau,\,\sigma),\,S_{\beta}(\tau,\,\sigma')\}
&=& \frac{i}{2}\del_{\al\beta}\,\del(\sigma - \sigma')
\,, \nn \\
\Biggl(
\{\psi_{\al}(\tau,\,\sigma),\, \psi^{\s T}_{\beta}
(\tau,\,\sigma')\} 
&=&  \frac{1}{2}\del_{\al\beta}\,\del(\sigma - \sigma')\,
\Biggr)\,, \nn
\end{eqnarray}
where the delta function is defined by 
\[
 \del(\sigma - \sigma') \;=\; \frac{1}{2\pi}
\sum_{n}\e^{in(\sigma - \sigma')}\,.
\]
From the above results, we can obtain the 
commutation relations of bosonic modes as 
\begin{eqnarray} 
\label{com-b}
\left[\bar{\al}^i_m, \, \al_n^j\right] 
&=& [\al^i_m,\,\bar{\al}^j_n] \;=\; 0 \,, \quad (i,j = 1,\ldots, 8)\,,
\\  
\left[\al_m^a,\, \alpha_n^{a'}\right] &=& 
\omega_m^{\s\rm B}\del_{m+n,0}\,\del^{aa'} \,,\quad (a,\,a' = 1,2,3,4)\,, \nn \\
\left[\bar{\al}_m^a,\,\bar{\al}_n^{a'}\right] 
&=& \omega_m^{\s\rm B}\del_{m+n,0}\,\del^{aa'}\,, \nn \\
 \left[\al_m^b,\, \alpha_n^{b'}\right] &=& 
\omega_m^{\s\rm B'}\del_{m+n,0}\,\del^{bb'} \,, \quad (b,\,b' = 5,6,7,8)\,, 
\nn \\ 
\left[\bar{\al}_m^b,\,\bar{\al}_n^{b'}\right] 
&=& \omega_m^{\s\rm B'}\del_{m+n,0}\,\del^{bb'}\,,  
\nn \\ 
\left[x_0^i,\,p^j_0\right] 
&=& i\del^{ij}\,,\quad \mbox{otherwise is zero.} \nn 
\end{eqnarray}
and those of fermionic ones as 
\begin{eqnarray}
\label{com-f}
\{(\Psi_m)_{\al},\, (\widetilde{\Psi}_n)_{\beta}^{\s T}
\} &=& 
\{
(\widetilde{\Psi}_m)_{\al},\,(\Psi_n)_{\beta}^{\s T}
\} \; = \; 0\,,  \\
\{(\Psi_m)_{\al},\,(\Psi_n)_{\beta}^{\s T}\}
&=& 
\{(\widetilde{\Psi}_m)_{\al},\,(\widetilde{\Psi}_n)_{\beta}^{\s T} 
\} \;=\;  \frac{1}{2}\del_{m+n,0}\,\del_{\al\beta}\,, \nn \\
\{(\Psi_m')_{\al},\, (\widetilde{\Psi}_n')_{\beta}^{\s T}
\} &=& 
\{
(\widetilde{\Psi}_m')_{\al},\,(\Psi_n')_{\beta}^{\s T}
\} \; = \; 0\,, \nn \\
\{(\Psi_m')_{\al},\,(\Psi_n')_{\beta}^{\s T}\}
&=& \{(\widetilde{\Psi}_m')_{\al},\,(\widetilde{\Psi}_n')_{\beta}^{\s T} 
\} \;=\;  \frac{1}{2}\del_{m+n,0}\,\del_{\al\beta}\,. \nn 
\end{eqnarray}
Though further considerations will not be done here, 
we can obtain the quantum Hamiltonian or spectrum 
exactly with the standard procedure.

\subsection{Open Strings and D$p$-branes in Type IIA on PP-wave}

In this subsection we shall discuss the mode-expansions of 
open strings in the type IIA string by imposing boundary conditions. 
In particular, we would like to 
consider D-branes, following Ref.\,\cite{DP}. (For more detailed studies,
see Refs.\,\cite{BP,BGG}.) It has been shown in Ref.\,\cite{DP} 
that D$p$-brane is not allowed for $p=1$, 9 and there are  
some restrictions on directions of allowed D-branes.      
First we consider the open string action described by 
\begin{eqnarray}
\label{open}
 \mathcal{S}_{st} &=& \frac{1}{2\pi}\int\! d\tau\!\! \int^{\pi}_0\!\!\!
d\sigma\, \mathcal{L}_{\rm st}\,, \\
  \mathcal{L}_{st} &=& \frac{1}{2\al'}\Biggl[
\sum_{i=1}^8\partial_+ x^i\partial_- x^i - 
\left(\frac{\mu}{3}\right)^2 \sum_{a=1}^4 (x^a)^2 
- \left(\frac{\mu}{6}\right)^2\sum_{b=5}^8(x^b)^2 \Biggr]  \\ 
&& \quad 
+i \Psi^{1+}{}^{\s T}\partial_-\Psi^{1+} 
+ i\Psi^{1-}{}^{\s T}\partial_-\Psi^{1-} 
+ i\Psi^{2+}{}^{\s T}\partial_+\Psi^{2+} 
+ i\Psi^{2-}{}^{\s T}\partial_+\Psi^{2-} \nn \\
&& \quad - i\frac{\mu}{3}\Psi^{1-}\widetilde{\Pi}^{\s T}\Psi^{2+} 
- i\frac{\mu}{6}\Psi^{1+}{}^{\s T}\widetilde{\Pi}^{\s T}\Psi^{2-} 
+ i\frac{\mu}{6}\Psi^{2-}{}^{\s T}\widetilde{\Pi}\Psi^{1+} 
+ i\frac{\mu}{3}\Psi^{2+}{}^{\s T}\widetilde{\Pi}\Psi^{1-}\,. \nn
\end{eqnarray}
Similarly, we obtain equations of motion (\ref{em1})-(\ref{em6}) 
from the above action (\ref{open}).     
In order to solve the above equations of motion 
we have to impose the following boundary conditions on 
bosonic coordinates $x^i$'s ($i=1,2,\cdots ,8$), 
\begin{eqnarray}
 && {\rm Neumann}~:\quad \partial_{\sigma}x^{\overline{m}} \;=\; 0\,, \quad 
(\,\overline{m} \;=\; +,\,-,\, {\rm and~some~} p-1~{\rm coordinates})\,, 
\nn \\
 && {\rm Dirichlet}~:\quad \partial_{\tau}x^{\underline{m}} \;=\; 
0\,,\quad 
(\,\underline{m} \;=\; {\rm other}~ 9-p ~{\rm coordinates})\,, \nn 
\end{eqnarray}
where 8 transverse indices $i=1,\,\ldots,\, 8$ are decomposed into 
$a = 1,\,2,\,3,\,4$~(flux directions) 
and $b = 5,\,\ldots,\,8$.  
For fermionic coordinates, boundary conditions are imposed as
\begin{eqnarray}
 \Psi^{1-}|_{\sigma =0,\,\pi} &=& 
\widetilde{\Omega}\Psi^{2+}|_{\sigma=0,\,\pi}\,,
\nn \\
 \Psi^{2+}|_{\sigma=0,\,\pi} &=& 
\widetilde{\Omega}^{\s T}\Psi^{1-}|_{\sigma=0,\,\pi}\,, \nn \\
 \Psi^{1+}|_{\sigma =0,\,\pi} &=& 
\widetilde{\Omega}\Psi^{2-}|_{\sigma=0,\,\pi}\,,
\nn \\
 \Psi^{2-}|_{\sigma=0,\,\pi} &=& \widetilde{\Omega}^{\s T}
\Psi^{1+}|_{\sigma=0,\,\pi}\,, \nn 
\end{eqnarray}
where $\widetilde{\Omega}$ is defined by 
\begin{eqnarray}
 \widetilde{\Omega} \;=\; \tg_{\underline{m}_1}\tg_{\underline{m}_2}
\ldots \tg_{\underline{m}_{9-p}}\,. \nn 
\end{eqnarray}
The $\widetilde{\Omega}$ includes odd number of gamma matrices
since the $SO(8)$ chiralities of $\Psi^1$ and
$\Psi^2$ must be opposite in the type IIA theory and hence 
$p$ is restricted to even. 

Under these boundary conditions 
we can obtain classical solutions for
equations of motion, 
and mode-expansions of bosonic variables are given by 
\begin{eqnarray}
x^{a}(\tau,\,\sigma) &=& x_0^{a} \cos\left(\frac{\mu}{3}\tau\right) 
+ \left(\frac{3}{\mu}\right) 2\al'p_0^{a}\sin\left(\frac{\mu}{3}\tau\right) 
\qquad (a=1,2,3,4)\nn \\
& & \qquad + i\sqrt{2\al'}\sum_{n\neq 0}\frac{1}{\omega_n^{\rm\s B}}
\,\alpha_n^{a}\,\e^{-i\omega_n^{\rm\s B}\tau}
\cos(n\sigma) 
\,, \quad ({\rm Neumann})
\\
x^{a}(\tau,\,\sigma) &=& \sqrt{2\al'}\sum_{n\neq 0}
\frac{1}{\omega_n^{\rm\s B}}
\,\alpha_n^{a}\, \e^{-i\omega_n^{\rm\s B}\tau}\sin(n\sigma)
\,, \quad ({\rm Dirichlet}) \\
x^{b}(\tau,\,\sigma) &=& x_0^{b} \cos\left(\frac{\mu}{6}\tau\right) 
+ \left(\frac{6}{\mu}\right)2\al'p_0^{b}\sin\left(\frac{\mu}{6}\tau\right) 
\quad (b=5,6,7,8)
\nn \\
&& \qquad 
+ i\sqrt{2\al'}\sum_{n\neq 0}\frac{1}{\omega_n^{\rm\s B'}}
\,\alpha_n^{b}\,\e^{-i \omega_n^{\rm\s B'}\tau}
\cos(n\sigma) 
\,, \qquad ({\rm Neumann})\\
x^{b}(\tau,\,\sigma) &=& \sqrt{2\al'}
\sum_{n\neq 0}\frac{1}{\omega_n^{\rm\s B'}}
\,\alpha_n^{b}\, \e^{-i \omega_n^{\rm\s B'}\tau}\sin(n\sigma)
\,, \quad ({\rm Dirichlet})
\end{eqnarray}
where $\omega_n^{B,\,B'}$ has been defined by (\ref{omega}). 
The mode-expansions of fermionic variables are the same as in the 
closed string case. 
The quantization can be done in the same way as closed strings. 
The commutation relations of bosonic and fermionic 
modes are the same as (\ref{com-b}) and (\ref{com-f}).  
The quantum Hamiltonian and 
spectrum can be also studied with the standard procedure 
but we will not investigate them furthermore here. 

Next we will study D-branes.
Though the mode-expansions of fermionic variables are the same as in the closed
string case, in the open string case fermionic boundary conditions lead to 
further constraints 
\begin{eqnarray}
&& \Psi_0 \;=\; \widetilde{\Omega}\widetilde{\Pi}\widetilde{\Psi}_0\,,\quad 
 \widetilde{\Omega}^{\s T}\widetilde{\Psi}_0 \;=\; - \widetilde{\Pi}\Psi_0\,,
\nn \\
&& \widetilde{\Psi}_n \;=\;\widetilde{\Omega}\Psi_n\,, \quad (n\neq 0) \,,\nn 
\label{bo2} \\
&& \Psi_0' \;=\; \widetilde{\Omega}\widetilde{\Pi}\widetilde{\Psi}_0'\,,\quad 
 \widetilde{\Omega}^{\s T}\widetilde{\Psi}_0' \;=\; 
- \widetilde{\Pi}\Psi_0'\,,
\nn \\
&& \widetilde{\Psi}_n' \;=\;\widetilde{\Omega}\Psi_n'\,, 
\quad (n\neq 0) \,.\nn 
\end{eqnarray}
The self-consistency of these conditions gives us a condition 
\begin{eqnarray}
 \widetilde{\Omega}\tPi\widetilde{\Omega}\tPi \;=\; -1\,. 
\end{eqnarray}
This condition is peculiar to the pp-wave, and gives an additional 
constraint for the D$p$-branes in the theory. In fact, in the massive
type IIB theory, D1- and D9-branes are forbidden and 
D3-, D5- and D7-branes can exist but those directions are limited. 
In the massive type IIA theory similar restrictions are imposed. 

We shall list the possible D$p$-branes below:
\begin{itemize}
 \item $p=8$~:\quad one of $I = 1,\,2,\,3$ is Dirichlet type, 
\[
 \widetilde{\Omega} = \tg^{\s I}\,,
\]
 \item $p=6$~:\quad 1)~ two of $I= 1,\,2,\,3$ and one of $I'=4,\,\ldots,\,8$
       are Dirichlet types, 
\[
 \widetilde{\Omega} = \tg^{\s I}\tg^{\s J}\tg^{\s I'}\,,
\] 
\qquad\qquad  2)~ three of $I'= 4,\,\ldots,\,8$ are Dirichlet types,
\[
 \widetilde{\Omega} = \tg^{\s I'}\tg^{\s J'}\tg^{\s K'}\,,
\] 
 \item $p=4$~:\quad 1)~ all of $I= 1,\,2,\,3$ and two of
       $I'=4,\,\ldots,\,8$ are Dirichlet types,
\[
 \widetilde{\Omega} = \tg^{\s I}\tg^{\s J}\tg^{\s K}\tg^{\s I'}\tg^{\s J'}\,,
\]
\qquad\qquad 2)~ one of $I= 1,\,2,\,3$ and four of $I'=4,\,\ldots,\,8$
       are Dirichlet types, 
\[
 \widetilde{\Omega} = \tg^{\s I}\tg^{\s I'}\tg^{\s J'}\tg^{\s K'}\tg^{\s L'}\,,
\]
 \item $p=2$~:\quad two of $I= 1,\,2,\,3$ and all of $I'=4,\,\ldots,\,8$ 
are Dirichlet types,
\[
 \widetilde{\Omega} = \tg^{\s I}\tg^{\s J}\tg^{\s I'}
\tg^{\s J'}\tg^{\s K'}\tg^{\s L'}\tg^{\s M'}\,.
\]
\end{itemize}    

In conclusion, all D$p$-branes can exist for $p=$~even, 
but those directions are 
constrained as the case of the type IIB \cite{DP}. 
We note that 
zero-point energy varies for each direction of D-branes.  

\section{Conclusions and Discussions}

We have considered the type IIA string theory on the pp-wave
background from the eleven-dimensional viewpoint. To begin, 
we have discussed the type IIA pp-wave solution through the toroidal
compactification of the maximally supersymmetric pp-wave solution in eleven
dimensions on a spatial isometry direction.
Next, we have derived the action of the type IIA string
theory from the type IIA pp-wave solution of the supergravity.
Moreover, we have derived the type IIA string action 
from the eleven-dimensional 
supermembrane theory on the maximally supersymmetric pp-wave 
background by applying  the double dimensional reduction for a 
spatial isometry direction. The resulting action 
agrees with the one 
obtained from the supergravity side. In particular, the Kaluza-Klein gauge
field induces a mass term of a 
bosonic coordinate in the type IIA theory. 
Furthermore, we have written down 
the action of the matrix string on the
pp-wave. This action contains the 3-point interaction and mass terms. 
Also, the field strength of the gauge connection 
is shifted. However, this action is still gauge invariant, though 
this theory is not maximally supersymmetric. In particular, this theory 
is reduced to 
the matrix string theory in the flat space 
by taking the limit $\mu \rightarrow 0$. 
We also discussed 
the quantization of closed and open strings in the type IIA string. 
In particular, the allowed D$p$-branes in this theory has been
investigated. The value $p=2,\,4,\,6$ and 8 are allowed but the directions 
of D-branes are constrained. 

We can also consider compactifications along 
other isometry directions. In such cases the number of the 
remaining supercharges is less than 24. 
It is nice to study the Type IIA pp-wave background 
preserving 26 supercharges \cite{26} 
or type IIA string theory from the 
eleven-dimensional supermembrane.    
It is an interesting work to discuss less supersymmetric type IIA 
string theories from the supermembrane. 
Moreover, the supersymmetric D-branes in such type IIA 
string theories are very interesting subject to study. 

It is nice to study the matrix string theory written down here 
from several aspects.   
In particular, it would be interesting to study the relation between 
the matrix string theory on the pp-wave and ``string bit'' \cite{Bit}.

\newpage 
  
\noindent 
{\bf\large Acknowledgement}

The work of K.S. is supported in part by the Grant-in-Aid from the 
Ministry of Education, Science, Sports and Culture of Japan 
($\sharp$ 14740115). 

\appendix 

\vspace*{1cm}
\noindent 
{\large\bf  Appendix}
\section{Compactification on an $SO(3)$-direction}
\label{app}

In the text we have considered the compactification on an
$SO(6)$-direction. We can also compactify one of the directions 
$x^{\s I}$'s, $({I}= 1,\,2,\,3)$. 
In this case, there are some different points from the 
compactification on an $SO(6)$-direction and we shall summarize 
these differences.  

In this compactification on an $SO(3)$-direction 
the change of variables is given by 
\begin{eqnarray}
&& X^+ \;=\; x^+\,,\quad X^- \;=\; x^- + \frac{\mu}{3}x^1x^2\,, \quad 
X^{a} \;=\; x^a\,,~(a = 3,\ldots,9)\,
\nn \\
&& X^1 \;=\; x^1 \cos\left(\frac{\mu}{3}x^+\right) 
- x^2\sin\left(\frac{\mu}{3}x^+\right)\,, 
\quad X^2 \;=\; x^1 \sin\left(\frac{\mu}{3}x^+\right) 
+ x^2 \cos\left(\frac{\mu}{3}x^+\right)\,,
\end{eqnarray} 
then the metric is rewritten as 
\begin{eqnarray}
 ds^2 &=& -2dx^+dx^- + G'_{++}(x^3,x^{\s I'})(dx^+)^2 
+ \sum_{r=1}^9(dx^r)^2 - \frac{4}{3}\mu x^2 dx^+ dx^1\,,  \\
&& G'_{++}(x^3,x^{\s I'}) \;\equiv\; -\left[
\left(\frac{\mu}{3}\right)^2(x^{3})^2 + \left(\frac{\mu}{6}\right)^2 
\sum_{{\s I'}=4}^9(x^{\s I'})^2 \right]\,, \nn
\end{eqnarray}
and the constant 4-form flux is still written in Eq.(\ref{flux}). 
In this case the $x^1$-direction is a manifest spatial isometry 
direction. In the same way, we can obtain the type IIA solution 
from the above expression. 
The ten-dimensional metric $g_{\mu\nu}$ is given by 
\begin{eqnarray}
g_{\mu\nu}dx^{\mu}dx^{\nu} &=& 
-2dx^{+}dx^- + g_{++}(x^i)(dx^{+})^2 + \sum_{i=1}^8(dx^i)^2\,,  \\
&& g_{++}(x^i) \;\equiv\; 
- \left[\left(\frac{2}{3}\mu\right)^2(x^{1})^2 
+ \left(\frac{\mu}{3}\right)^2(x^2)^2 
+ \left(\frac{\mu}{6}\right)^2 \sum_{a= 3}^8 (x^a)^2
\right]\,, \nn 
\end{eqnarray}
where we have relabelled the indices $x^2,\cdots,x^9$ as 
$x^1,\,\cdots,x^8$. 
The Kaluza-Klein gauge field $A_{\mu}$ is represented as
\begin{equation}
 A_{+} \;=\; - \frac{2}{3}\mu x^1\,,\quad A_{i} \;=\; 0\,,~(i=1,\ldots,8)\,,
\end{equation}
and non-zero NS-NS 2-form is given by 
\begin{eqnarray}
 B_{+1} \;=\; \frac{\mu}{3}x^2\,, \quad 
B_{+2} \;=\; -\frac{\mu}{3}x^1\,. 
\end{eqnarray}
In this case, the R-R 3-form $C_{\mu\nu\rho}$ and dilaton $\phi$ vanish.

To begin, let us discuss the bosonic sector. 
We can easily obtain the bosonic action from Eq.(\ref{nlsm}) 
using the type IIA metric. The resulting action is expressed as 
\begin{eqnarray}
 S_{\rm B} &=& \frac{1}{4\pi\al'}\int\!d\tau\!\!\int^{2\pi}_0\!\!\!d\sigma\, 
\Bigg[
\sum_{i=1}^8\left[
(\partial_{\tau}x^i)^2 - (\partial_{\sigma}x^i)^2
\right] + \frac{2}{3}\mu\, x^2\partial_{\sigma}x^1 
- \frac{2}{3}\mu\, x^1 \partial_{\sigma}x^2 \nn \\ 
&& \qquad 
- \left(\frac{2}{3}\mu\right)^2(x^1)^2 - \left(\frac{\mu}{3}\right)^2 (x^2)^2 
- \left(\frac{\mu}{6}\right)^2 \sum_{a=3}^8(x^a)^2 \Bigg]\,,
\label{so3}
\end{eqnarray}
where the mass term for $x^1$ is induced from the Kaluza-Klein gauge field
$A_{\mu}$ as the case of the compactification on an
$SO(6)$-direction. It is also an easy exercise 
to derive the above action (\ref{so3}) 
by using the double dimensional reduction. 

Next, we shall consider the fermionic sector. 
Now, in the study of the supergravity, 
the field strength of R-R 3-form is zero, but 
NS-NS 2-form is non-zero and it has the constant field strength 
proportional to $\mu$. Thus, this contribution induces the fermion mass
term. 
However, there might be possibly an issue for the numerical constant 
and the fermion mass term obtained in the supergravity analysis is 
not identical with the one derived via double dimensional reduction 
if we naively use the 
expression of the covariant derivative in the text.

\end{document}